\documentclass[twocolumn,showpacs,preprintnumbers,amsmath,amssymb]{revtex4-1}

\usepackage{dcolumn}
\usepackage{float}
\usepackage{graphicx}
\usepackage{siunitx}

\begin{document}

\title{Scattering-Free Optical Levitation of a Cavity Mirror}
\author{G. Guccione$^{\dagger,*}$, M. Hosseini$^{\dagger,*}$, S. Adlong$^{\dagger,*}$, M. T. Johnsson$^{*}$, J. Hope$^{*}$, B. C. Buchler$^{\dagger,*}$, P. K. Lam$^{\dagger,*}$}

\affiliation{$^{\dagger}$Centre for Quantum Computation and Communication Technology, $^{*}$Department of Quantum Science, The Australian National University, Canberra, Australia}

\date{\today}

\begin{abstract}
We demonstrate the feasibility of levitating a small mirror using only radiation pressure. In our scheme, the mirror is supported by a tripod where each leg of the tripod is a Fabry-Perot cavity. The macroscopic state of the mirror is coherently coupled to the supporting cavity modes allowing coherent interrogation and manipulation of the mirror motion. The proposed scheme is an extreme example of the optical spring, where a mechanical oscillator is isolated from the environment and its mechanical frequency and macroscopic state can be manipulated solely through optical fields. We model the stability of the system and find a three-dimensional lattice of trapping points where cavity resonances allow for build up of optical field sufficient to support the weight of the mirror. Our scheme offers a unique platform for studying quantum and classical optomechanics and can potentially be used for precision gravitational field sensing and quantum state generation.
\end{abstract}

\maketitle
Recently much effort has been directed toward the development of new fabrication methods and experimental techniques for controlling optomechanical interactions at the quantum level~\cite{Teufel:2011:Nature, Chan:2011:Nature}. Optomechanical effects have been observed in mechanical objects with masses ranging from femtograms, as in nano-optomechanical systems~\cite{Sun:2012:NanoLett}, to kilograms in the case of gravitational wave antennae~\cite{Corbitt:2006:PhysRevA:b}. Reaching the quantum regime in optomechanical systems is fundamentally interesting as one is then in a position to prepare macroscopic quantum states, which can, for example, be employed in tests of large-scale quantum decoherence~\cite{Bose:1999:PhysRevA} and models of gravity~\cite{Pikovski:2012:NatPhys, Yang:2013:PhysRevLett}. The main barrier to reaching the quantum regime is thermalization resulting from intrinsic coupling to environmental reservoirs. This is generally hard to avoid since most mechanical oscillators are supported by some mechanical structure that acts as a bridge for thermal fluctuations. One method to limit thermalization is to operate in cryogenic environments. Nevertheless, the dissipation of energy through the mechanical support still contributes significantly to the decoherence of the mechanical state~\cite{Hao:2003:SensActAPhys}. Fabrication of a phononic-band gap structure into the substrate~\cite{Safavi-Naeini:2010:OptExp} has been proposed as one way to reduce the dissipation. Optical trapping~\cite{Ni:2012:PhysRevLett} and levitation~\cite{Libbrecht:2004:PhysLettA, Arvanitaki:2013:PhysRevLett, Kiesel:2013:arXiv, Singh:2010:PhysRevLett} have also been suggested as possible routes to low-dissipation quantum optomechanics. In the recent proposals, despite the mechanical support being completely removed, scattering from the levitated object leads to interaction with the environment and lowering of optomechanical coupling.

Radiation pressure within an optical resonator can be used to couple the mechanical oscillations of a suspended cavity mirror with the optical mode~\cite{Groblacher:2009:NatPhys, Corbitt:2007:PhysRevLett}. Such coupling between optical and mechanical systems can be used for a variety of applications, including precision measurement~\cite{Rehbein:2008:PhysRevD}, the creation of a mechanical quantum harmonic oscillator~\cite{Groblacher:2009:NatPhys, Gigan:2006:Nature, Arcizet:2006:Nature, Schliesser:2006:PhysRevLett}, control of quantum macroscopic coherence~\cite{Mancini:1997:PhysRevA}, the generation of squeezed light for quantum information~\cite{Marino:2010:PhysRevLett, Chang:2010:PNAS}, optomechanical entanglement between an oscillator and a cavity field~\cite{Vitali:2007:PhysRevLett}, and also reversible mapping of quantum states of light into mechanical excitations~\cite{Fiore:2011:PhysRevLett}. Current optomechanical experiments rely on radiation pressure to drive a mirror on a mechanical spring~\cite{Groblacher:2009:NatPhys, Corbitt:2007:PhysRevLett}. In such systems, the coherence time of mechanical oscillations is limited by clamping losses and thermalization.

In this paper we propose a unique approach toward this problem, wherein the material supports are completely eliminated. We consider a vertical geometry where the upper cavity mirror floats on the radiation pressure exerted by intra-cavity fields. This system constitutes an extreme example of environmental isolation because the motion of the centre of mass is naturally decoupled from the internal degrees of freedom in addition to being mechanically isolated by levitation~\cite{Ashkin:1976:ApplPhysLett}. This approach provides an elegant route toward the elimination of the mirror clamping losses, and will allow high mechanical quality factors and the potential to reach the quantum regime of phonon-photon interaction.

An optically suspended mirror will provide a highly configurable platform for performing a variety of experiments in cavity optomechanics. Not only will the mechanical oscillator be isolated from environmental noise, scattering noise, and mechanical losses, but the spring constant and damping coefficient will be selectable through choice of optical frequency and power.
\begin{figure}[!t]
	\centerline{\includegraphics[width=0.7\columnwidth]{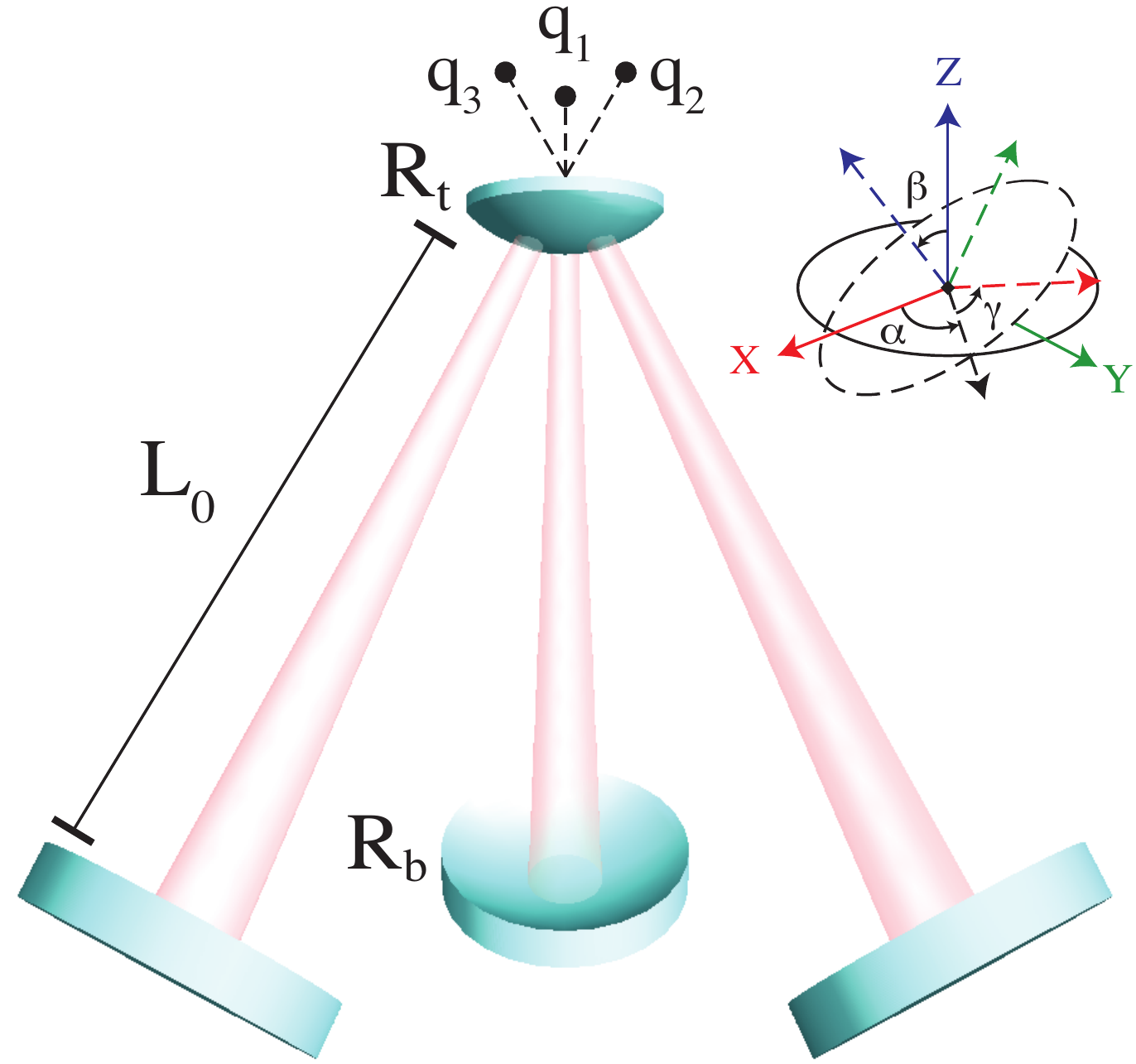}}
	\caption[]{Arrangement for the tripod optical cavity with the convex end mirror levitated on the three optical springs. The three lower mirrors are identical; $\mathbf{q}_1$, $\mathbf{q}_2$ and $\mathbf{q}_3$ are the centres of curvature of the three lower mirrors; cavity lengths $L_0$ allowed by this configuration for optical stability are between $17$ and $20$~cm.}
	\label{fig: tripod}
\end{figure}

\textit{Optical forces and stability.}\ --- The optical spring effect has been observed in various systems~\cite{Hossein-Zadeh:2007:OptExp, Sheard:2004:PhysRevA, Fan:2009:ApplPhysLett} where the measured mechanical resonant frequency ($\omega_m$) depends on the optical power. We propose to create an optical spring, not just to provide additional rigidity to a weak mechanical spring, but to actually support a cavity mirror using radiation pressure alone. Stable suspension of the mirror can be attained through the use of a tripod-beam configuration shown in Fig.~\ref{fig: tripod}. Each beam will be the fundamental mode of a high-finesse cavity formed between the levitated mirror and one of the fixed lower mirrors. Appropriate use of active or passive damping, combined with the optical spring effect, will stabilize the suspended mirror on the optical fields.

The optical spring mirror that we propose is a convex mirror with radius of curvature of $R_t = 3$~cm and diameter of $2$~mm that is coated with high reflective (HR) dielectric materials and has reflectivity of $99.98-99.998\%$. Such high-reflective coatings typically have a laser damage threshold \SI{ \approx 30}~{MW/mm}$^2$, much greater than the intensity as anticipate on the mirror. The mirror substrate is made out of fused silica and has a mass around $0.3$~mg. The three lower mirrors are HR coated with $99.8-99.98\%$ reflectivity and have radius of curvature $R_b = 20$~cm. The cavity decay rate is given by $\kappa = \pi c / (\mathcal{F} L_0)$, where $L_0$ is the mean length of the cavity, and $\mathcal{F}$ is the cavity finesse.

The full position and orientation of the upper mirror is defined by the position of its centre of curvature $\mathbf{r}$, which we write in Cartesian coordinates $\{x, y, z\}$, and the \textsc{z-x-z} Euler angles $\{\alpha, \beta, \gamma\}$ that define its orientation from the canonical position. The orientations of the three lower mirrors are defined by the position of their centres of curvature $\mathbf{q}_n$, where $n = 1, 2, 3$ refers to the three cavities. The optical cavities form between the centre of curvature of the upper mirror and the centres of curvature of the lower mirrors, with lengths $L_n = R_b - R_t + \left\| \mathbf{q}_n - \mathbf{r} \right\|$. The laser power $P_n$ inside each cavity is given by $P_n = P_n^{(in)} \mathcal{F} / [1 + \mathcal{F}^2 \sin^2\!{(k L_n)} ]$, where $k = 2 \pi / \lambda$, $\lambda$ is the wavelength and $P_n^{(in)}$ is the input power of the laser driving that cavity. This circulating power translates into a force $\mathbf{F}_n$ on the mirror with magnitude $F_n = 2 P_n / c$. A total power of approximately $3$~W in the three cavity beams combined, a near-paraxial geometry, and cavity finesse of $1000$ will give a force sufficient to suspend the mirror.
\begin{figure}[t]
	\centerline{\includegraphics[width=\columnwidth]{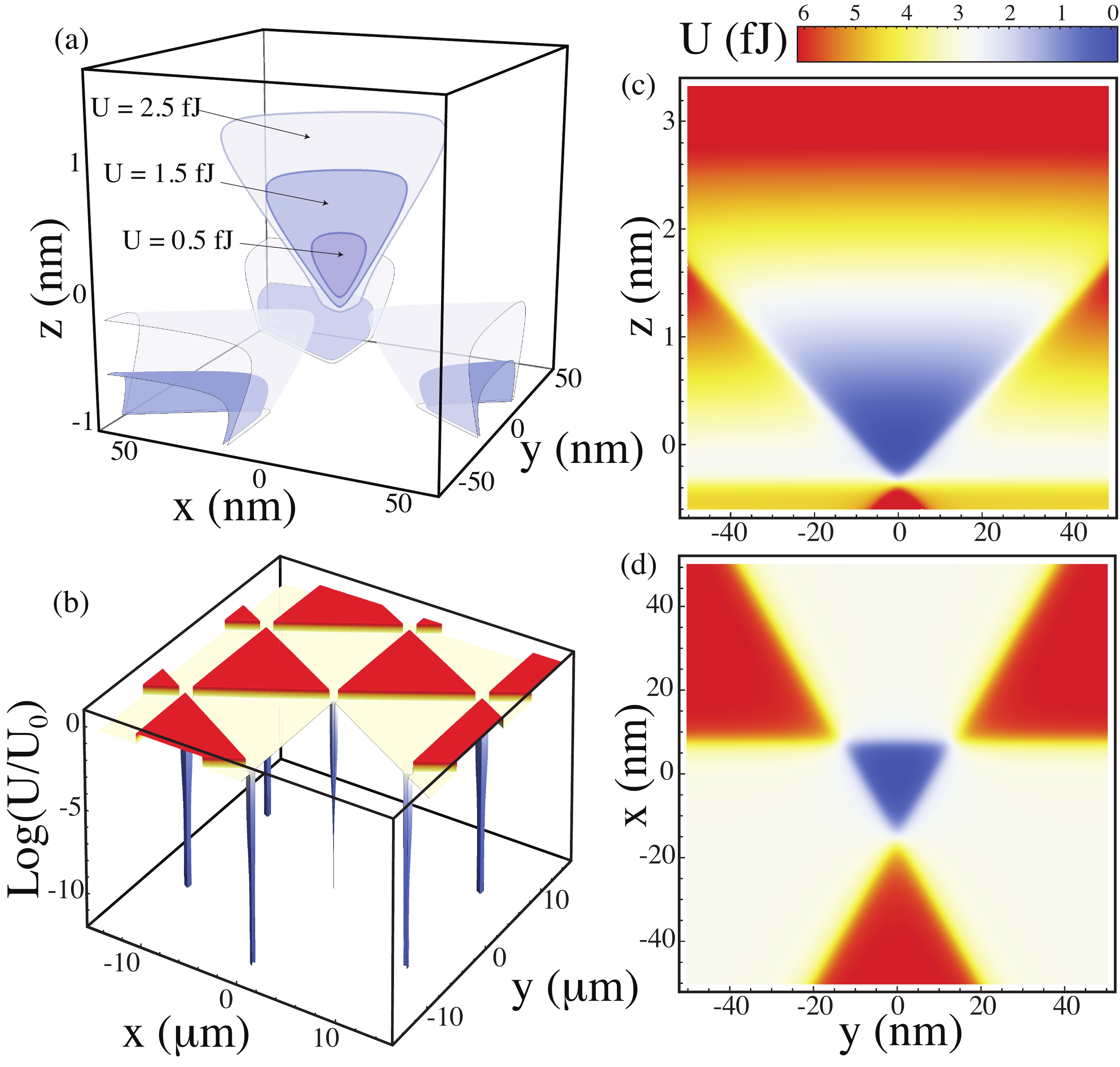}}
	\caption[]{{\bfseries (a)} Isopotential surfaces showing the stability region in space. {\bfseries (b)} Triangular lattice of trap sites showing the trapping potential of the mirror on \textsc{x-y} with trapping sites spaced from each other by approximately $15$~$\mu$m. The potential is in logarithmic scale, normalized to its value just outside of the traps, $U_0$. {\bfseries (c)-(d)} Trapping potential on \textsc{y-z} and \textsc{x-y} around the tight-confinement region. For these plots a finesse of $1000$, a total input power of $3$~W, and a mirror mass of $0.3$~mg were used.}
	\label{fig: potential}
\end{figure}
When the mean radiation pressure force cancels the gravitational pull on the mirror any variation in the intra-cavity power can produce a damping or restoring force, depending on the cavity field detuning. If we consider a case where each cavity field is blue-detuned from the resonance condition, any shortening of the cavity will result in an increase of power and therefore of the radiation pressure force, and lengthening of the cavity will result in a decrease of the force. This suggests that there will be a restoring force allowing the floating mirror to be stable for small fluctuations.

\begin{figure}[t]
	\centerline{\includegraphics[width=\columnwidth]{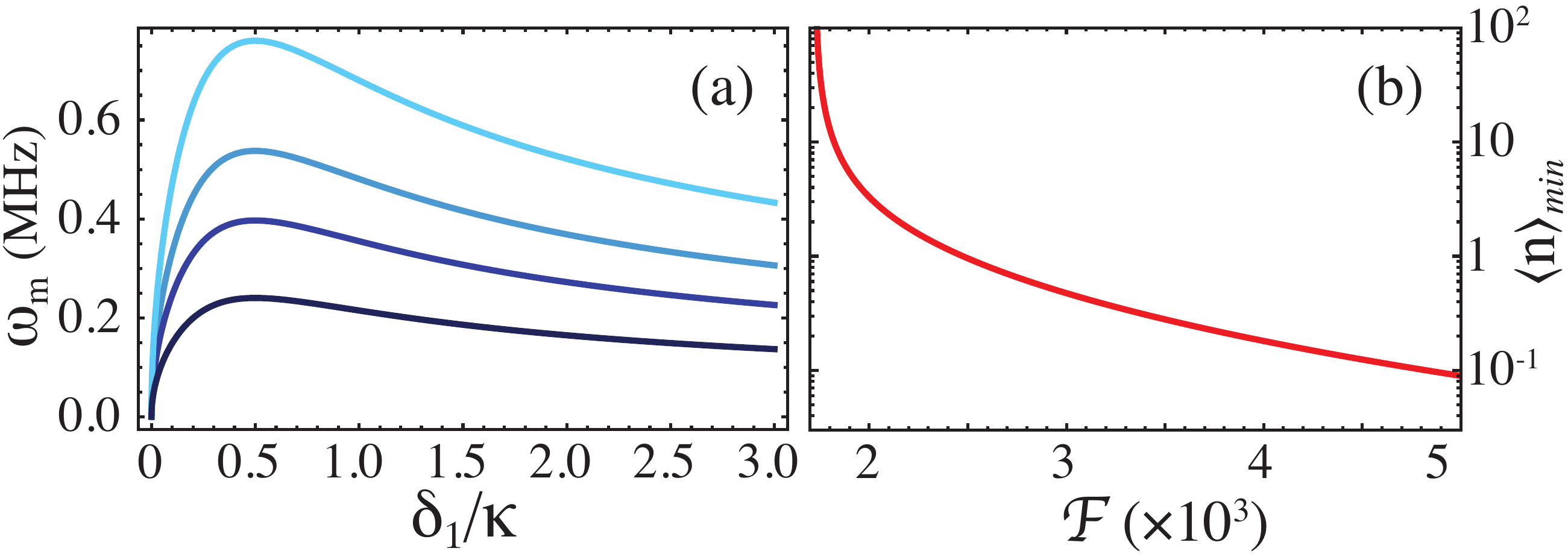}}
	\caption[]{{\bfseries (a)} Plot of the mechanical frequency of the mirror on the optical spring versus cavity detuning of the trapping beam normalized to the cavity linewidth for four different cavity finesses (from darkest to lightest, $\approx 1000$, $3000$, $5000$ and $10000$). {\bfseries (b)} Minimum mean phonon number plotted as a function of cavity finesse for a trapping beam of detuning $\delta_1 = 0.5 \kappa$ ten times weaker than the cooling beam detuned by $\delta_2 = -\omega_m$.}
	\label{fig: detuning}
\end{figure}

The mechanical stability is best analyzed by constructing a generalized potential $U(\mathbf{r}, \alpha, \beta, \gamma)$ for the six coordinates describing the top mirror. This potential is independent of $\alpha$ and $\gamma$, and trivially stable with respect to $\beta$. Setting $\beta=0$, it is given by
\begin{eqnarray}
	\label{eq: potential}
	U = \sum_{n = 1}^3 \frac{2 P_n^{(in)} }{c} \frac{\tan^{-1}\! \left[ \mathcal{F} \, \tan\! \left(k L_n(\mathbf{r}) \right) \right]}{k} + m g z	.	\nonumber
\end{eqnarray}
For displacements of the top mirror there is a large, three-dimensional lattice of similar tight-confinement spots. The potential near the trapping sites is visualized in Fig.~\ref{fig: potential}, where we see isopotential surfaces in (a), showing that the stable region can be up to $30$~nm wide in the horizontal directions, and approximately $1$~nm wide along the $\vec{z}$-axis. In the $\textsc{xy}$-plane there is a triangular lattice of trap sites approximately $15$~$\mu$m apart, as shown in part (b). In parts (c) and (d) we see 2-D sections of the potential near a central trapping site. The potential depths of these trapping sites scale almost linearly with the finesse and input power of the cavities, however increasing the finesse also reduces the spatial size of the traps in each dimension. Tuning the frequency of the three beams to find the trapping sites could prove very effective, as a very small fractional changes in the laser frequency of each cavity enables the entire lattice to be scanned. Optical stability is obtained when the mirror is precisely positioned at one of the equilibrium points, corresponding to a cavity detuning of $\kappa / 2$, and the cavity length will then be self-locked by the radiation pressure gradient. In principle, no active feedback is required to keep the cavity stable, although it may nevertheless be desirable. The optical spring stiffness tensor can be calculated by taking the second derivative of the work done on the mirror~\cite{Sidles:2006:PhysLettA}, and from it the mechanical frequency can be determined. Fig.~\ref{fig: detuning}(a) shows the $\vec{z}$-axis frequency for four choices of finesse, as a function of trapping beam detuning.

While the blue-detuned field (trapping beam of detuning $\delta_1$) will create a strong trapping force, it will also result in some anti-damping force on the mirror~\cite{Corbitt:2006:PhysRevA:b, Kippenberg:2007:OptExp}. To counteract this we use a second, red-detuned field with detuning $\delta_2$. When the cavity linewidth is less than the mechanical resonance frequency of the mirror ($\kappa \ll \omega_m$), a laser tuned to the red motional sideband of the cavity will amplify the scattering of light into the main cavity mode, thus removing energy from the mirror~\cite{Wilson-Rae:2007:PhysRevLett, Marquardt:2007:PhysRevLett, Schliesser:2009:NatPhys}. The detuning of the cooling beam needs to be equal to the mechanical frequency of the mirror, $\delta_2 = - \omega_m$, which depends on finesse as well as the detuning of the trapping beam from the cavity resonance.

%\begin{eqnarray}
%	\label{eq: stiffness}
%	\chi_{i j} = - \frac{2}{c} \sum_{n = 1}^{3}{(P_n^{(in)} \partial_{\mathbf{r}_{i},\mathbf{r}_{j}} L_n + \partial_{\mathbf{r}_{i}} P_n^{(in)} \partial_{\mathbf{r}_{j}} L_n)}	\nonumber
%\end{eqnarray}
%where indices $i, j = 1,2,3$ represent the three coordinates $\{x,y,z\}$.

\textit{Background gas collisions.}\ --- Background gas collisions with the mirror can increase or decrease the mechanical energy of the mirror depending on its size. Assuming that the mirror operates in the free molecular flow regime~\cite{Bhiladvala:2004:PhysRevE}, the mechanical dissipation rate due to fluid friction is given by $\gamma_m = 2 \rho_g v_g S/m$, where $m$ is the mass of the mirror, $S$ its cross-section, $\rho_g$ is the density of gas and $v_{g} = \sqrt{2 k_B T / m_{g}}$ is the mean of the magnitude of the velocity of a gas molecule of mass $m_g$ at temperature $T$ in any one dimension. At a pressure $P$ of $10^{-8}$~bar, a dissipation rate of about $10^{-5}$~Hz is estimated, which suggests a mechanical Q factor of about $5 \times 10^{9}$ for a mechanical frequency of $500$~kHz. Since the mechanical Q of the levitated mirror is not limited by intrinsic mechanical dissipation, lowering the pressure will linearly enhance the Q factor.

Assuming a gas molecule undergoing an elastic collision with the mirror, the collision rate can be written as $\Gamma_g(v_g) = P S v_g / (k_B T)$. The energy dissipation rate of the mirror can be calculated by
\begin{eqnarray}
	\label{eq: gamma_bg}
	\frac{dE_g}{dt} = \int_0^{\infty}{\!\!\!dv \; \Gamma_g(v) D(v) \frac{2m_g^2}{m}v^2}
\end{eqnarray}
where $D(v)$ is the Maxwell-Boltzman velocity distribution. Neglecting the dissipation and noise sources due to blackbody and laser power fluctuations (described below), one can estimate the thermal phonon numbers $\langle n_{th} \rangle =\dot E_g / (\gamma_m \hbar \omega_m)$ to be around $50$ for a vacuum pressure of $10^{-8}$ bars. This is already a low initial phonon number occupation that can be further reduced by laser cooling.

\textit{Laser noise.}\ --- Laser intensity noise causes fluctuation of the optical spring constant. To determine the anti-damping rate arising because of this we follow a similar method as taken in the context of trapping atoms in optical traps~\cite{Gehm:1998:PhysRevA,Savard:1997:PhysRevA}.

Fluctuations in the intra-cavity photon number alter the mechanical frequency of the trap. We shall focus on stochastic fluctuations of the optical spring stiffness. The dominant parametric heating rate, $R_{n \to m}$, arises from the component of the noise power spectrum at the second harmonic. The rate of transition for the cavity mirror from a state with $n$ phonons to a state with $m \neq n$ phonons during a time period of $\tau$ is only nonzero when $m = n \pm 2$ and can be simplified as $R_{n \to n \pm 2} = \frac{\pi \omega_m^2}{16} S_\epsilon(2 \omega_m) (n + 1 \pm 1) (n \pm 1)$
where $S_\epsilon(\omega)=2/\pi\int_0^{\infty}{dt' \cos{\omega t'}\langle \epsilon(t)\rangle\langle\epsilon(t+t')\rangle}$ is the one-sided power spectrum of the fractional fluctuation, and $\epsilon(t)$ is fractional intensity noise.

We can see that the shot noise leads to parametric transitions (where the phonon number $n \to n \pm 2$ jumps in pairs) at a rate proportional to the power spectral density of the fluctuations at frequency $2 \omega_m$. The heating rate due to intensity fluctuations is given by
\begin{eqnarray}
	\label{eq: gamma_I}
	\gamma_I = \frac{\sum_n{P_n (R_{n \to n + 2} - R_{n \to n - 2})}}{\sum_n{P_n \hbar \omega_m (n+1/2)}} = \frac{\omega_m^2}{4} S_\epsilon(2 \omega_m)
\end{eqnarray}
where $P_n$ is the probability that the mirror occupies a state with $n$ phonons. The average energy increases exponentially with an $e$-folding time of $\tau_e = \gamma_I^{-1}$. Assuming a mirror oscillation frequency of $500$~kHz, an energy $e$-folding time of $10$~s requires $S_\epsilon\simeq 1.3 \times 10^{-6}$~$Hz^{-1}$. Hence, if most of the intensity noise were evenly distributed over a bandwidth of $300$~kHz, the root-mean-square fractional intensity noise of the laser should be less than $(\int_0^{\infty}{d\omega S_\epsilon(\omega)})^{1/2}= 7 \times 10^{-4}$.

\textit{Blackbody radiation.}\ --- A small fraction of the light incident on the mirror will be absorbed due to its finite absorption coefficient, and with no means of mechanical dissipation in vacuum the only way to dissipate this energy is through blackbody radiation~\cite{Rybicki:1979:WileyVCH}. A fraction of the absorbed light results in increasing the internal temperature of the mirror, $T_{int}$.

The internal heating rate due to blackbody absorption is given by $dE/dt = \sum{\hbar c k R_{abs}(k)}$, where the sum is over all blackbody radiation modes (and polarizations), $k$ is the wavevector of each mode, and the total energy absorption rate is given by:
\begin{eqnarray}
	\label{eq: absorption rate}
	\frac{dE_{abs} }{dt} = \frac{S}{4 \pi^2} \int_0^{\infty}{\!\!\!dk \; \hbar c^2 k^3 n_k \epsilon_{b}}=\frac{\pi^2 S~ \epsilon_{b} (k_B T)^4}{60~ c^2 \hbar^3}
\end{eqnarray}
where $n_k = 1/(e^{\hbar c k / k_B T} - 1)$ is the probability occupation of each mode, $\epsilon_b$ is the temperature-independent blackbody emissivity of the mirror. The blackbody emission rate is then given by $-dE_{abs}/dt$ where $T\to T_{int}$. Taking into account the blackbody absorption and emission as well as laser absorption heating, a temperature raise of $\Delta T_b < 1$~K can be inferred for $\alpha = 10^{-5}$~m$^{-1}$ and $\epsilon_b \approx 2 \times 10^{-4}$~\cite{Fontana:1988:ApplOpt}. We note that the net work done by blackbody radiation on the mirror over one oscillation is zero due to the time-independent nature of the radiation.

\textit{Optical cooling.}\ --- Now we investigate the possibility of cooling the mirror close to its quantum ground state even when starting from room temperature. It has been shown that for sufficiently high mechanical frequencies cooling in cryogenic devices is sufficient~\cite{OConnell:2010:Nature}, although the ground state can also be achieved using laser cooling~\cite{Chan:2011:Nature}.

In the resolved sideband regime a laser field red detuned from the cavity resonance frequency by the mechanical frequency will result in cooling of the motion of the mirror. This is because the cavity will enhance process whereby a phonon from the mirror is added to the photon, giving light that is resonant with the cavity mode. We can add such a laser field to our system, however the cooling achieved is limited by the heating due to the trapping beam, which is blue-detuned from cavity resonance. Denoting the trapping and cooling beams as $\lambda=\{1,2\}$ respectively, we can write the net laser cooling rate of the mirror due to both intra-cavity fields as~\cite{Wilson-Rae:2007:PhysRevLett, Marquardt:2007:PhysRevLett}, $\gamma_{rp} = G^2 \sum_{\lambda = 1,2}[S_\lambda(-\omega_m) - S_\lambda(+\omega_m)]$,
where $G = \omega_c \sqrt{\hbar / (2 \, m \, \omega_m)}/L_0$ is the optomechanical coupling, $\omega_c$ is the cavity resonance frequency, $S_\lambda(\omega) = \bar{n}_\lambda \kappa / [(\kappa/2)^2 + (\omega + \delta_\lambda)^2]$ is the power spectrum of the laser noise, and $\bar{n}_\lambda$ is the mean photon number of the optical field. Ignoring all other sources of damping, one can write an expression for the mean thermal phonon occupation:
\begin{eqnarray}
	\label{eq: phonon number}
	\frac{{\langle n \rangle}_{min} + 1}{{\langle n \rangle}_{min}} = \frac{S_1(+\omega_m) + S_2(+\omega_m)}{S_1(-\omega_m) + S_2(-\omega_m)}
 \end{eqnarray}
In a typical optomechanical system, the minimum phonon number attained by laser cooling is ${\langle n \rangle}_{min}\approx (\kappa/ 4\omega_m)^2$, limited only by the cooling beam~\cite{Marquardt:2007:PhysRevLett}. This lower bound is a result of back action of the cooling beam on the mirror that is equivalent to Doppler cooling limit in atomic systems~\cite{Wineland:1979:PhysRevA}. In our scheme, the trapping beam limits the cooling process, and since the mechanical frequency depends on detuning and power of the trapping beam, laser cooling becomes a bigger challenge. We find that ground state cooling can be achieved provided the cavity finesse is larger than $3000$ and detunings of trapping and cooling beams from cavity resonance are respectively: $\delta_1 \approx \kappa/2$ and $\delta_2 = -\omega_m$. A plot of minimum mean phonon number at the optimal detunings is shown in Fig.~\ref{fig: detuning}(b) as a function of cavity finesse.

Both laser intensity fluctuations and background collisions are mechanisms of damping that lower the effective mechanical Q of the mirror. Assuming we are in the regime of negligible laser noise, the coupling to a thermal reservoir increases the attainable mean phonon number by $\langle N \rangle = (\gamma_{rp} {\langle n \rangle}_{min} + \gamma_m \langle n_{th}\rangle)/(\gamma_{rp} + \gamma_m)$. A high finesse cavity at low vacuum pressures offers a very low mechanical dissipation $\gamma_m$ and minimum phonon number $\langle n \rangle_{min}$ well below one, hence ground state of the mirror can, in principle, be reached by cavity cooling.

\textit{Applications.}\ --- The proposed optomechanical system can provide ultra-low-dissipation mechanical vibration and large optomechanical coupling suitable for various purposes. We briefly consider two possible applications: gravitational measurements and squeezing.

The optical spring means that any change in weight of the mirror will linearly alter the intra-cavity and output optical power. The gravitational acceleration $g$ will therefore be linear with the cavity output power. Assuming a shot-noise limited laser and impedance-matched cavities, we find $\delta g / g = \delta P / P = 1 / \sqrt{n_{ph}}$ where $n_{ph}$ is the mean photon number. Detecting $100$~mW of power thus gives a precision around $10^{-11}$ for an integration time of $100$~s. This level of performance complements, and could present sensitivity improvements to, modern atom interferometry techniques~\cite{Peters:2001:Metrologia}.

Due to the nonlinear interaction between the intra-cavity field and the mirror position, squeezing at very low frequency can be achieved~\cite{Brooks:2012:Nature, Marino:2010:PhysRevLett}. Mechanical and optical squeezing can also be accomplished by adding a sinusoidally varying component to the intensity of the trapping beam, inducing parametric amplification of the amplitude and phase quadratures of the motion. This type of squeezing has been demonstrated by modulating the qubit nanoresonators, where a gain of $30$~dB and thermal noise squeezing of $4$~dB was achieved~\cite{Suh:2010:NanoLett}.

In conclusion, we devised an optomechanical system in which a cavity mirror can be suspended and be maximally decoupled from the environment on three optical springs. The proposed system suppresses the scattering-induced heating and clamping dissipation and is an ultimate example of optical levitation. We showed that such system provides an isolated macroscopic oscillator with a very high mechanical quality factor. We also investigated the possibility of reaching quantum regime by means of laser cooling.

This research was funded by the Australian Research Council Centre of Excellence (CE110001027) and Discovery Program (DP1092891) schemes.

\vfill

\end{document}